\documentclass[prl,twocolumn,aps,a4paper,showpacs,amsfonts,amsmath,amssymb]{revtex4-1}

\usepackage{graphicx}

\begin{document}

\title{Volatility polarization of non-specialized investors' heterogeneous activity}
\author{Mario Guti\'errez-Roig} 
\author{Josep Perell\'o}
\affiliation{Departament de F\'{\i}sica Fonamental, Universitat de Barcelona.\\ Mart\'i i Franqu\'es 1, E-08028 Barcelona, Spain}

\begin{abstract}
Financial markets provide an ideal frame for studying decision making in crowded environments. Both the amount and accuracy of the data allows to apply tools and concepts coming from physics that studies collective and emergent phenomena or self-organised and highly heteregeneous systems. We analyse the activity of $29\,930$ non-expert individuals that represent a small portion of the whole market trading volume. The very heterogeneous activity of individuals obeys a Zipf's law, while synchronization network properties unveil a community structure. We thus correlate individual activity with the most eminent macroscopic signal in financial markets, that is volatility, and quantify how individuals are clearly polarized by volatility. The assortativity by attributes of our synchronization networks also indicates that individuals look at the volatility rather than imitate directly each other thus providing an interesting interpretation of herding phenomena in human activity. The results can also improve agent-based models since they provide direct estimation of the agent's parameters.
\end{abstract}

\date{\today}


\maketitle

Collective behavior in socioeconomic contexts is becoming more and more empirically studied from mathematical and physical point of view since large amount of data is now available. This leads to the emergence of a new data-driven research area \cite{king}. Most intrigued aspects in this area concern how microscopic interactions trigger macroscopic phenomena and how individuals react to such current macroscopic bath \cite{schelling}, showing some analogy with magnetism and Ising Model \cite{krawiecki,krause,preis}. Nowadays, human activity leaves a digital trace that can be correlated to global information flows. Both the amount and accuracy of the data makes financial trading floors to be ideal scenarios to study the linkage between collective and individual human activity. Recent research has related aggregated market trading volume with search queries in Google or Yahoo considered as a macroscopic field \cite{preis,bordino,da}, but individual activity data is not easily accessible for research purposes. Just a few papers have been published with this sort of data \cite{Lillo,Grinblatt2009,Lachapelle,Tuminello} and none of them focuses on the relationship between microscopic and macroscopic levels. In this regard, the main purpose of the paper is to answer the question about how do non-expert investors make decisions and whether their actions are a response to a given macroscopic field with rather unique data from an Spanish investment firm. Since the volume traded by these investors represents a very small subset of the whole volume of the market participants, we can therefore observe the influence or polarization of macro-level signals over the microscale, but not the other way around \cite{krawiecki,krause}. Moreover, the conclusions can be easily extrapolated to the broadest context of human decision making since we deal with non-expert agents.  

Every individual decision is materialised in an operation, either buying or selling, and each investor trades with his/her own money. We here analyse a dataset that contains $3\,303\,695$ individual recordings from $29\,930$ clients of a particular investment firm. Price, date and, number of shares traded from each transaction were stored on a daily basis. Individuals were trading between 2000 and 2007 in 120 assets of the Spanish stock market, IBEX. During this period the market had no general trend. We limit our analysis to the 8 most traded assets by our investors: Telefonica (TEF), Ezentis (EZE), Sogecable (SGC), Santander (SAN), BBVA, Red Electrica (ELE), Repsol (REP) and Zeltia (ZEL), that belong to different economic sectors in order to seek universality. As we focus on human activity, we filter the automatic operations still retaining $84.5 \%$ of the TEF data (worst) and $99.2 \%$ of the SGC data (best). Table~\ref{tab:tab1} summarizes the data subset.

\begin{figure}
\begin{flushleft}
A)
\end{flushleft} 
\includegraphics{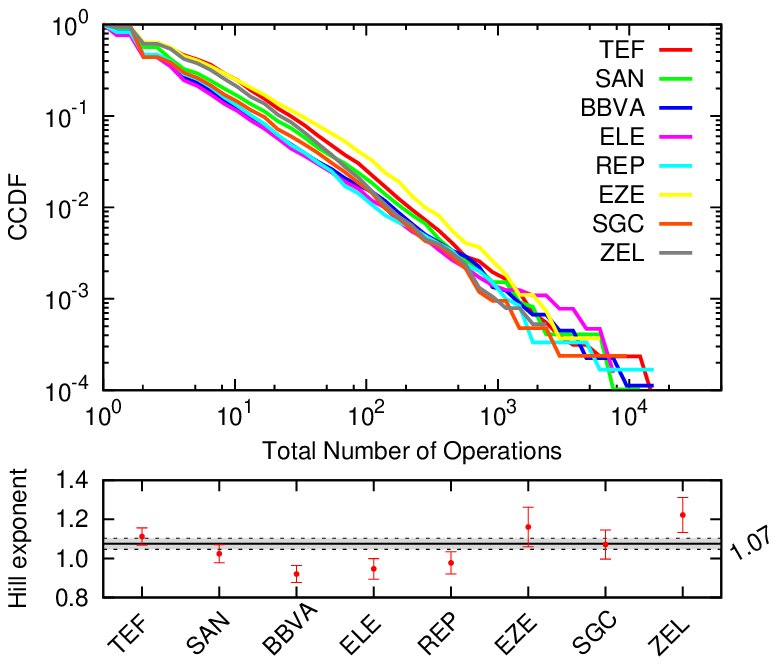}
\begin{flushleft}
B)
\end{flushleft} 
\includegraphics{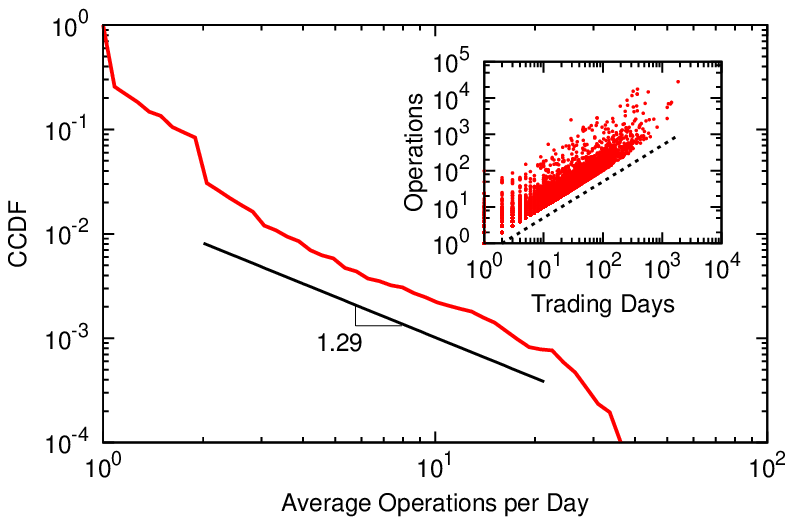}
\caption{ Activity properties. A) Activity CCDF of individuals in the 8 assets (Top). The Hill tail indexes together with the index obtained from data aggregation, $1.07\pm0.03$ (Down). B) The aggregate of the 8 assets activity CCDF of individuals as a function of the number of operations per trading day. The inset represents the number of operations versus trading days jointly with a dotted line of unit slope.}
\label{fig:fig1}
\end{figure}

It is known that human activity is bursty and non-homogeneous in time and several descriptions have been proposed \cite{Barabasi}. Figure~\ref{fig:fig1}A displays the complementary cumulative distribution function (CCDF) of individuals' activity showing a robust power-law with an exponent very close to 1 (Zipf's law), both for the aggregated data and for each stock separately. The exponent coincides with the one found in several contexts \cite{Barabasi,Vazquez,Blasius} and particularly for Nokia expert traders \cite{Tuminello}. It can however be argued that the heterogeneous activity profile is simply due to the fact that the time period between the first and the last operation is different among investors and thus we cannot infer a different decision-making profile for each individual. Inset of Fig.~\ref{fig:fig1}B partially supports this argument indeed, since there is a linear relation between the number of operations and the number of trading days, even though data points are widely scattered. Therefore, the number of operations has been normalised and the distribution of the number of operations per trading day (OpD) has been represented in Fig.~\ref{fig:fig1}B. From main Fig.~\ref{fig:fig1}B we conclude that heterogeneity is still preserved with a tail index that equals 1.29. Those individuals that operate very infrequently, less than 1 operation per day, represent around $75\%$ of the population, while there is an investor with around 30 operations per trading day on average. 

\begin{table}
\begin{tabular}{lrrr}
\hline
\hline
Ticker &\centering Individuals & \centering Operations & Trading Volume\\
\hline 
TEF & \centering $11571$ &\centering $273,150$ & $91.62\times 10^{9}$\\
SAN & \centering $9450$ &\centering $159,798$ & $100.24\times 10^{9}$\\
BBVA & \centering $8549$ &\centering $128,006$ & $38.77\times 10^{9}$\\
ELE &\centering $6226$ &\centering $89,452$ & $13.87\times 10^{9}$\\
REP &\centering $5655$ &\centering $81,045$ & $12.06\times 10^{9}$\\
EZE &\centering $2696$ &\centering $60,421$ & $2.37\times 10^{9}$\\
SGC &\centering $4182$ &\centering $57,816$ & $1.56\times 10^{9}$\\
ZEL &\centering $3694$ &\centering $52,601$ & $1.21\times 10^{9}$\\
\hline
\hline
\end{tabular}
\caption{Total number of clients, number of operations and trading volume (in euros) for studied assets.}
\label{tab:tab1}
\end{table}

\begin{figure}
\begin{flushleft}
A)
\end{flushleft} 
\includegraphics[width=7cm]{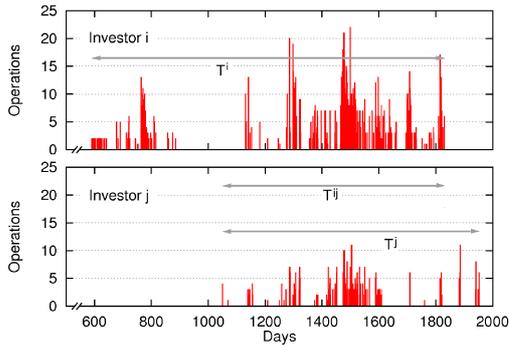}
\begin{flushleft}
B)
\end{flushleft} 
\includegraphics[width=7cm]{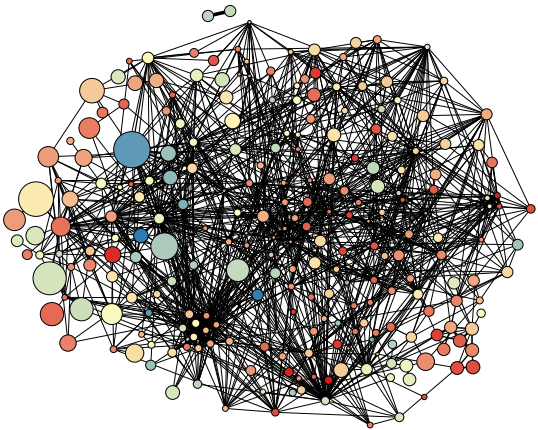}
\caption{A) Activity profile of two investors ($i$ and $j$). $T^{ij}$ is the intersection between active periods $T^{i}$ and $T^{j}$. B) Synchronization network for REP. The size of the nodes is proportional to OpD. The node's colour represents $\rho_{O\nu}^i$ from $-1$ (red) to $+1$ (blue) and edge's thickness the weight $\rho^{ij}$.}
\label{fig:fig2}
\end{figure}

We define the activity $O(t)^i$ of investor $i$ as the number of operations performed at day $t$. We then understand $T^{i}$ as the time period starting the day that $i$ did the first operation and ending when $i$ did the last operation. We note that the active period $T^{i}$ may also include non-trading days, so the $N^i$ total number of trading days of $i$ follows inequality $N^{i}\leq T^i$. As shown in Fig.~\ref{fig:fig2}A, we compute cross-correlation between investors $i$ and $j$ with simultaneous activity during a period $T^{ij}>0$. That is:
\begin{equation}
\label{rho_ij}
\rho^{ij}= \displaystyle{\frac{1}{T^{ij}}}\displaystyle{ \sum_{t=T_{F}^{ij}}^{T_{L}^{ij}}\displaystyle{\frac{\left(O(t)^{i}-\bar{O}^i\right)(O(t)^{j}-\bar{O}^j)}{\sigma_{O^i}\sigma_{O^j}}}},
\end{equation}
where $T_F^{ij}$ and $T_{L}^{ij}$ are respectively first day and last day of $i$ and $j$ simultaneous activity. The bar represents the averages and the $\sigma$'s denote the standard deviations of the $i$ and $j$ investor's activity over $T^{ij}$. Identical notation will be used in forthcoming equations. In order to work with statistical robustness, we limit to those individuals with at least 20 operations and focus on the most synchronized ones. To keep the most synchronized ones, we shuffle the values of $O(t)^{i}$ and $O(t)^{j}$ belonging to the time period $T^{ij}$ and calculate cross-correlation again. If the original $\rho^{ij}$ is below the threshold corresponding to the $0.01$ p-value for all shuffled terms, we set $\rho^{ij}$ to zero manually. This double filtering still maintains most of the operations and volume traded, as shown in Tab.~\ref{tab:tab2}. Figure~\ref{fig:fig2}B shows the Repsol synchronisation network as an illustration. The nodes (individuals) without connections are removed in the interest of clarity and weighted edges correspond to the filtered coefficients $\rho^{ij}$. The rest of stocks are reported in Tab.~\ref{tab:tab2}. We also report modularity of each stock which gives an idea of the network structure. Their values are similar to other social and biological studied networks \cite{Newman2006}. These magnitudes tell us that agents' activity is far from randomness and unveils some community structure. 

Interesting questions arising here are which relevant factors lead individuals to make decisions and if agents are influenced by macroscopic information. Recent research has correlated Google searches with trading volume \cite{preis,da,bordino} but here we aim to identify an endogenous variable at an individual level. In this sense, the most relevant macroscopic variable in terms of the market activity is volatility –rather than price \cite{podobnik}. We work with the High-Low volatility $\nu(t)$ defined as the difference between the highest and the lowest price value divided by the open price of day $t$ \cite{Bouchaud}. The easiest way to see how volatility influences our investors’ activity is by focussing on mesoscopic activity variables from the same day: $O(t)$, that is the total number of operations made by the studied individuals. The dependences between $O(t)$ and $\nu(t)$ time-series are studied by computing linear cross-correlations (Long). Due to the fact that volatility is a long memory process \cite{Bouchaud}, the mean in the correlation formula is subtracted to avoid any bias. However, most of the studied investors have a short-term horizon. To consider this, we compute as well the correlation by subtracting the local mean with a 5-day Moving Average (Short) \cite{Bouchaud}. Table~\ref{tab:tab2} shows significant correlation in both measures. So high volatility clearly affect clients' activity to operate both in long and short time horizons.

\begin{table}
\begin{tabular}{lcccccccccccc}
\hline
\hline
Ticker & Investors & Operations & Mod. & \multicolumn{2}{c}{$\rho_{O\nu}$} \\
&&&& Long & Short \\
\hline
TEF & \centering $1240(10\%)$ &\centering $204\,146(74.74\%)$ & $0.411$ &$0.5560$ & $0.5259$\\
SAN & \centering $701(7\%)$ &\centering $114\,537(71.68\%)$ &$0.385$ &$0.2795$ & $0.5259$\\
BBVA & \centering $385(4\%)$ &\centering $90\,719(70.87\%)$ &$0.468$ &$0.0858$ & $0.4175$\\
ELE &\centering $257(4\%)$ &\centering $64\,107(71.67\%)$ &$0.463$ &$0.2399$ & $0.3983$\\
REP &\centering $252(4\%)$ &\centering $53\,542(66.06\%)$ &$0.531$ &$0.0972$ & $0.1504$\\
EZE &\centering $251(9\%)$ &\centering $43\,717(72.35\%)$ &$0.265$ &$0.2139$ & $0.3507$\\
SGC &\centering $135(3\%)$ &\centering $33\,959(58.74\%)$ &$0.473$ &$0.4789$ & $0.4684$\\
ZEL &\centering $251(7\%)$ &\centering $30\,611(58.19\%)$ &$0.423$ &$0.4228$ & $0.4893$\\
\hline
\hline
\end{tabular}
\caption{Number of investors with their number of operations in the activity synchronization network after removing inactive and non-synchronized investors from the original data (remaining percentage in brackets). Modularity calculated by Louvain method (Mod.) is provided. Last columns show the correlation at mesoscopic level between $O(t)$ and $\nu(t)$.}
\label{tab:tab2}
\end{table}

\begin{figure}
\includegraphics{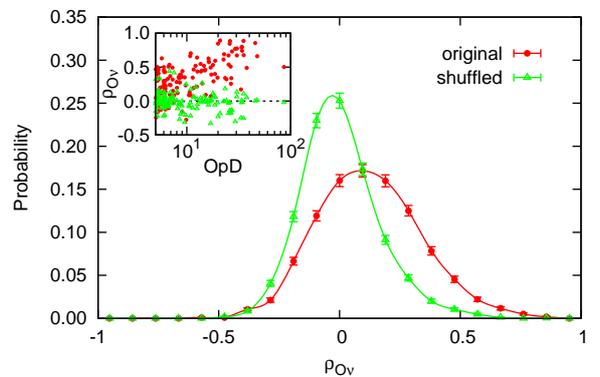}
\caption{Distribution of investors according its $\rho_{O\nu}$ for the aggregate of the eight assets. Inset shows a scatter plot with all investors cross-correlation with original and shuffled data.}
\label{fig:fig3}
\end{figure}

After analysing the collective response to volatility, the individual response needs to be regarded. If the investor $i$ is sensitive to risk fluctuations, activity $O(t)^i$ will not be time-homogeneous. We thus compute the correlation between volatility and number of operations
\begin{equation}
\rho_{O\nu}^{i}= \displaystyle{\frac{1}{N^{i}}}\displaystyle{ \sum_{t=T_{F}^{i}}^{T_{L}^{i}}\displaystyle{\frac{\left(O(t)^{i}-\bar{O}^i\right) \left(\nu(t)-\bar{\nu}\right)}{\sigma_{O^i}\sigma_{\nu}}}}.
\label{rho_Ov}
\end{equation}
The correlation only takes into account trading days ($O(t)^{i}>0$). Once more, investors with less than $20$ trading days are excluded. Figure~\ref{fig:fig3} shows that the population distributed according to its individual response to volatility $\rho_{O\nu}^{i}$ is not neutral nor symmetric. In all the studied stocks, the mean and the most probable value are shifted systematically to the positive side explaining the mesoscopic polarization reported in Tab.~\ref{tab:tab2}. The variance of the distribution is $1.8$ times the variance of the shuffled case that decorrelates volatility and activity time series. Therefore the population of agents in terms of $\rho_{O\nu}^i$ is sensitive to the studied macroscopic signal. The average shows an alignment to volatility and the variance increases, demonstrating a rather diverse response to volatility. Moreover, the most active daily investors display a special sensitivity to daily volatility as can be seen inset of Fig.~\ref{fig:fig3}. 

\begin{figure}
\includegraphics{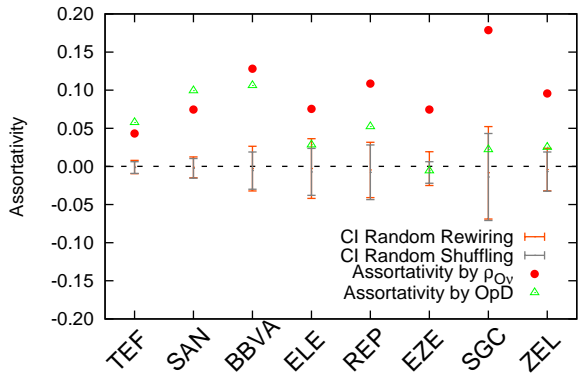}
\caption{Assortativity for all networks. Rewiring (green) and shuffling (blue) benchmark assortativities are shown. The error-bars provide $95\%$ confidence interval (CI).}
\label{fig:fig4}
\end{figure}

We can also guess that the activity profile of two agents with similar response to volatility should not be very different. The nodes colour in the synchronization network in Fig.~\ref{fig:fig2}B gives an idea of the cross-correlation $\rho_{O\nu}^i$ for each individual and a previous statement can intuitively be confirmed. We observe there that investors with similar response to volatility tend to be connected. We can go further with this idea in a more rigorous way by measuring the connectivity structure of a network through the assortativity by attributes according to the $\rho_{O\nu}^i$ value~\cite{Newman}. To do so, we first discretize all $\rho_{O\nu}^i$ by keeping the integer of $\rho_{O\nu}^i \times 100$. Final score for the assortativity is calculated as
\begin{equation}
r=\frac{1}{\sigma_{a} \sigma_{b}}\sum_{xy} xy(e_{xy}-a_x a_y),
\label{assort_eq}
\end{equation}
where $e_{xy}$ is the fraction of the links that join together nodes with values $x$ and $y$ for the discretized $\rho_{O\nu}^i$. The variables $a_x=\sum_{y} e_{xy}$ and $b_x=\sum_{x} e_{xy}$ are respectively the fraction of edges that start and end at nodes with values $x$ and $y$, and condition $\sum_{xy} e_{xy}=1$ needs to be fulfilled. The assortativity values in all 8 stocks lie between $0.08$ and $0.20$ as shown in Figure~\ref{fig:fig4}. Although these values are not very high, they are absolutely significant since they are clearly over the confidence intervals of two random benchmarks. First, randomized network rewires all the edges preserving node's properties. Second, one preserves the network's topology but shuffle node's attributes $\rho_{O\nu}^i$. In both cases, we keep population distribution and assortativity falls to 0. Similarly we perform the same analysis but taking into account the OpD property, instead. In this case, investors mostly synchronize in typical speculative assets (TEF, SAN, BBVA and REP).

This paper highlights and quantifies the influence of macro-variables in individuals' activity at a micro-level. The empirical work deals with rather unique records from a large set of non-expert investors (clients of a firm) from the Spanish stock market and they allow us to make statistically robust statements. The fact of the analyzed individuals being non-professional makes it possible to extrapolate the results obtained to other contexts with the purpose of better understanding human activity sensitive to a common macroscopic signal \cite{krawiecki,krause}. We have first observed that the activity is strongly heterogeneous among the individuals with a power-law exponent close to $1$. The influence of a macroscopic signal, the volatility, has been revealed both at a mesoscopic level, all investor's community, and at a microscopic level, each individual. The general tendency of individuals to show positive $\rho_{O\nu}^i$ explains the positive correlation at mesoscopic levels. The analysis of the assortativity by $\rho_{O\nu}^i$ over the activity synchronisation network has finally shown that the synchronisation among individuals takes place because investors are influenced in the same way by the same macroscopic field. Providing, thus, an alternative explanation for herding behaviour. The results obtained can also be very useful to test and calibrate or to even improve existing agent-based models \cite{krawiecki,Chiarella}. And most importantly they enhance the debate on rationality and decision-making in socioeconomic contexts based on physicist's perspective.

\begin{acknowledgments}
Financial support from Direcci\'on General de Investigaci\'on under contract FIS2009-09689 is acknowledged.
\end{acknowledgments}

\end{document}